\documentclass[onefignum,onetabnum]{siamonline220329}
\usepackage{amssymb}
\usepackage{amsmath}
\usepackage{comment}
\usepackage[utf8]{inputenc}
\usepackage{url} 
\usepackage{lipsum}
\usepackage{amsfonts}
\usepackage{graphicx}
\usepackage{epstopdf}
\usepackage[linesnumbered,ruled,vlined,algo2e]{algorithm2e} % typeset algorithms

\ifpdf
  \DeclareGraphicsExtensions{.eps,.pdf,.png,.jpg}
\else
  \DeclareGraphicsExtensions{.eps}
\fi

\usepackage{enumitem}
\usepackage[capitalize]{cleveref}
\usepackage{tikz}
\usetikzlibrary {arrows.meta}

\newcommand{\dT}{\overrightarrow{T}}
\newcommand{\T}{T}
\newcommand{\N}{\mathcal{N}}
\newcommand{\M}{\mathcal{M}}
\newcommand{\MM}{P}
\setlist[enumerate]{leftmargin=.5in}
\setlist[itemize]{leftmargin=.5in}

\newsiamremark{remark}{Remark}
\newsiamremark{hypothesis}{Hypothesis}
\crefname{hypothesis}{Hypothesis}{Hypotheses}
\newsiamthm{claim}{Claim}
\newsiamthm{Aproblem*}{Assignment Problem}
\newsiamthm{Oproblem*}{Orientation Problem}

\headers{Extending Robinson Spacess}{F. Brucker, P. Préa, and C. Thraves Caro}

\title{Extending Robinson Spaces: Complexity and Algorithmic Solutions for Non-Symmetric Dissimilarity Spaces\thanks{Part of this work was done while C. Thraves Caro was visiting the École Centrale Méditerranée}}
\author{Fran\c cois Brucker\thanks{LIS, Aix-Marseille Université, CNRS, Université de Toulon, Marseille, France, and École Centrale Méditerranée, Marseille, France. 
  (\email{francois.brucker@lis-lab.fr}, \email{pascal.prea@lis-lab.fr}).}
\and Pascal Préa\footnotemark[2]
\and Christopher Thraves Caro\thanks{Departamento de Ingenier\'ia Matem\'atica, Facultad de Ciencias F\'isicas y Matem\'aticas, Universidad de Concepci\'on, Chile (\email{cthraves@udec.cl}).}
}
\usepackage{amsopn}

\ifpdf
\hypersetup{
  pdftitle={Extending Robinson Spaces: Complexity and Algorithmic Solutions for Non-Symmetric Dissimilarity Spaces},
  pdfauthor={F. Brucker, P. Préa, and C. Thraves Caro}
}
\fi

\begin{document}

\maketitle

% REQUIRED
\begin{abstract}
In this work, we extend the concept of Robinson spaces to asymmetric dissimilarities, enhancing their applicability in representing and analyzing complex data. Within this generalized framework, we introduce two different problems that extend the classical seriation problem: an optimization problem and a decision problem. We establish that these problems are NP-hard and NP-complete, respectively. Despite this complexity results, we identify several non-trivial instances where these problems can be solved in polynomial time, providing valuable insights into their tractability. 
\end{abstract}

% REQUIRED

% REQUIRED

\section{Introduction}

Similarity and dissimilarity measures are fundamental to various data analysis and mining methodologies. Traditionally, these measures are assumed to be symmetric, a property that simplifies their application but limits their scope. Recent studies have increasingly highlighted the importance of accommodating non-symmetric measures, which more accurately model real-world scenarios where relationships are inherently directional \cite{amigo2020foundations, 7424392, 8332621}. In this work, we focus on extending the concept of Robinson spaces to encompass these asymmetric measures, broadening the applicability of this mathematical framework in complex data environments.
\subsection{Context and generalities}
Let \(X\) be a set of \(n\) elements. A \textit{dissimilarity} on \(X\) is (typically) a symmetric function \(d\) from \(X\times X\) to the nonnegative real numbers such that \(d(x, y) = 0\) if \(x = y\). The value \(d(x, y)\) represents the dissimilarity between \(x\) and \(y\), and \((X, d)\) is a \textit{dissimilarity space}. A total order \(<\) on \(X\) is \textit{compatible} if for all three elements \(x, y, z\) in \(X\) such that \(x < y < z\), it holds that \(d(x, z) \geq \max\{d(x, y), d(y, z)\}\). Due to the symmetry of $d$, the reverse order is also compatible if $<$ is compatible. A dissimilarity space is \textit{Robinson} if it admits a compatible order. 
Robinson dissimilarities were invented to order archaeological deposits~\cite{robinson_1951} chronologically and are now a standard tool for the seriation problem (see~\cite{prea2014optimal} for examples of application). Moreover, they are equivalent to pyramids, a classical model for classification~\cite{Did,DuFi} and helpful in recognizing tractable cases for the TSP problem~\cite{CeDeWo}.
In this document, we aim to extend the concept of Robinson spaces by considering non-symmetric dissimilarities. 
%To do so, we begin with the following analysis.

Any total order on \(X\) equates with a graph, specifically a path. Given a total order \(p_1< p_2< \cdots< p_n\) on \(X\), we construct the path \(P = (X, E)\), where \(E\) is the set \(\{\{p_i, p_{i+1}\}\colon i \in \{1, 2, \ldots, n-1\}\}\). Conversely, if we have a path \(P = (X, E)\), %whose set of vertices is \(X\), 
we derive a total order on \(X\) by traversing \(P\) from one end to the other, ordering its vertices according to their appearance. Using this equivalence, we redefine Robinson spaces as follows. A dissimilarity space \((X, d)\) is Robinson if there is a path \(P = (X, E)\) such that \(E = \{\{p_i, p_{i+1}\colon i \in \{1, 2, \ldots, n-1\}\}\), and for every three vertices \(p_i, p_j, p_k\) with \(i < j < k\), it holds that \(d(p_i, p_k) \geq \max\{d(p_i, p_j), d(p_j, p_k)\}\). In this case, we say that $P$ is Robinson. 

Brucker in \cite{brucker2005hypertrees} extended the notion of Robinson spaces to more complex structures, extending the definition given in the last paragraph. A dissimilarity space \((X, d)\) is \textit{Tree-Robinson} (or \textit{T-Robinson}) if there is a tree \(T = (X, E)\) such that for every pair of elements \(x, y\) in \(X\), the \(\{x, y\}\)-path in \(T\) is Robinson.

Determining whether a dissimilarity space \((X, d)\) is Robinson or T-Robinson is equivalent to finding a graph \(T = (V, E)\) —precisely, a path or a tree, respectively— and a bijection \(B: X \rightarrow V\) such that for every pair of elements \(x, y\) in \(X\), the \(\{B(x), B(y)\}\)-path in \(T\) is Robinson.

We extend the concept of dissimilarity spaces by considering \emph{non-symmetric} dissimilarities. Furthermore, we use directed graphs to define \emph{non-symmetric} T-Robinson spaces. From now on, we focus on dissimilarity spaces that are not necessarily symmetric. 

%\begin{definition} 
A dissimilarity space \((X, d)\) is \emph{one-way-Robinson} if there exists a directed path \(\overrightarrow{P} = (X, \overrightarrow{E})\), where \( \overrightarrow{E}=\{(p_i,p_{i+1})\colon i \in \{1, 2, \ldots, n-1\}\}\), such that for any three vertices \(p_i, p_j, p_k\) with \(i < j < k\), it holds that \(d(p_i, p_k) \geq \max\{d(p_i, p_j), d(p_j, p_k)\}\). In that case, we say that $\overrightarrow{P}$ is \emph{one-way-Robinson}, and the order $p_1<p_2<\ldots<p_n$ is said to be {\em compatible}.
%\end{definition}

It is important to note that the reverse orientation of \(\overrightarrow{P}\) is not necessarily Robinson since \(d\) is not necessarily symmetric. Therefore, we introduce a stronger definition.
%\begin{definition}

A dissimilarity space \((X, d)\) is \emph{two-way-Robinson} if there exists a directed path \(\overrightarrow{P} = (X, \overrightarrow{E})\), where \( \overrightarrow{E}=\{(p_i,p_{i+1})\colon i \in \{1, 2, \ldots, n-1\}\}\), such that for any three vertices \(p_i, p_j, p_k\) with \(i < j < k\), it holds that \(d(p_i, p_k) \geq \max\{d(p_i, p_j), d(p_j, p_k)\}\) and 
$d(p_k, p_i) \geq \max\{d(p_k, p_j), d(p_j, p_i)\}$. In that case, we say that $\overrightarrow{P}$ is \emph{two-way-Robinson}, and the order $p_1<p_2<\ldots<p_n$ is said to be {\em compatible}.
%\end{definition}

We aim to understand the complexity and provide algorithmic solutions for the following two problems.
\begin{Aproblem*}
Given a dissimilarity space \((X, d)\) and a directed tree \(\dT = (V, \overrightarrow{E})\) with \(|X| = |V|\), determine if there is a bijection between \(X\) and \(V\) such that every directed path in \(\dT\) is one-way-Robinson.
\end{Aproblem*}   

A second variation aims to recover the best possible orientation. Given a dissimilarity space \((X, d)\) and a tree \(T = (X, E)\), it is always possible to orient the edges of \(T\) such that every directed path in \(\dT\) is one-way-Robinson. This can be achieved by orienting the edges of \(T\) so that no directed path has more than one edge. Since all dissimilarity spaces with two elements are Robinson, every directed path will be one-way-Robinson. Therefore, the problem is:
\begin{Oproblem*}
        Given a dissimilarity space \((X, d)\) and an undirected tree \(T = (X, E)\), find an orientation \(\overrightarrow{T}\) of \(T\) that maximizes the number of pairs of elements \(x, y\) in \(X\) such that the directed \(\{x, y\}\)-path in $\dT$ is one-way-Robinson.
\end{Oproblem*}

In this document, we give an $O(n^3)$ algorithm to recognize two-way-Robinson dissimilarities (Theorem \ref{thm:two-way-rob}), establish that the Orientation Problem is NP-Hard (Theorem \ref{THEO_Orientation_NP}), and prove that the Assignment Problem remains NP-Complete even when the tree $T$ is a simple path (Theorem \ref{CORO_Assignment_NP}). Additionally, we identify a set of non-trivial instances where the Orientation Problem can be solved efficiently and provide optimal algorithms for these cases. %(Theorem \ref{Theo_optimal_orientation}).

The remainder of this document is structured as follows. Section \ref{sec:two-way-rob} presents an \( O(n^3) \) algorithm for recognizing two-way-Robinson spaces, where \( n \) is the size of the space. In Section \ref{sec:nphard-Or-As}, we prove that the Orientation problem is NP-Hard and the Assignation problem is NP-Complete. Section \ref{sec:optimalAllPathRob} focuses on cases where all paths in the input tree are Robinson, demonstrating that the Orientation problem can be solved optimally in such instances. Finally, Section \ref{SECTION_tract_cases_non_constant} identifies two additional cases in which the Orientation problem can be solved optimally: first, when the space is symmetric and the input tree is a star, and second, when the input tree is a path. For the first case, we also extend the technique to provide a polynomial-time algorithm for the Assignment problem.
We end this introduction with a short review of related works and some preliminaries.

\subsection{Related Work}

W. S. Robinson first defined Robinson spaces in \cite{robinson_1951} during his study on the chronological ordering of archaeological deposits. The \textit{Seriation} problem introduced in his work aims to determine whether a dissimilarity space is Robinson and, if so, to find a compatible order.

Numerous researchers have explored the recognition of Robinson spaces. Mirkin et al. presented an $O(n^4)$ recognition algorithm in \cite{graphsandgenes84}, where $n$ is the size of \(X\). Chepoi et al., utilizing divide and conquer techniques, introduced an $O(n^3)$ recognition algorithm in \cite{chepoi1997recognition}. Pr\'ea and Fortin later provided an optimal $O(n^2)$ recognition algorithm using PQ trees in \cite{prea2014optimal}. Laurent and Seminaroti, in \cite{laurent2017lex}, leveraged the relationship between Robinson spaces and unit interval graphs presented in \cite{roberts69} to introduce a recognition algorithm using Lex-BFS with a time complexity of $O(Ln^2)$, where \(L\) is the number of distinct values in the dissimilarity function. In \cite{carmona2024modules}, Carmona et al. used \textit{mmodules} and \textit{copoint partitions} to design a simple divide and conquer algorithm for recognizing Robinson spaces in optimal \(O(n^2)\) time.

In contrast to classical Seriation, the recognition of T-Robinson spaces has garnered less attention, with the primary contribution being Brucker's development of a recognition algorithm with a time complexity of \(O(n^5)\) in \cite{brucker2005hypertrees}. The circular variant has been explored by Armstrong et al. \cite{armstrong2021optimal} and Carmona et al. \cite{carmona2023simple}, who proposed two optimal algorithms for recognizing \emph{strict-circular-Robinson} spaces, each employing distinct techniques.

\subsection{Preliminaries}

Given a tree $T=(V,E)$, an \emph{orientation} $\overrightarrow{E}$ of $T$ is an assignment of either $(x,y)$, that we denote $x \rightarrow y$, or $(y,x)$, that we denote $y \rightarrow x$, for each $\{x,y\}\in E$. 
For $x, y \in V$, we will denote by $x\rightsquigarrow y$ if there is a directed path from $x$ to $y$, by $x \not\rightsquigarrow y$ if there is no directed path from $x$ to $y$, by $x\leftrightsquigarrow y$ if $x\rightsquigarrow y$ or $y\rightsquigarrow x$ and by $x\not\leftrightsquigarrow y$ if $x \not\rightsquigarrow y$ and $y\not\rightsquigarrow x$.
Notice that, for every $x\in V$, we have $x\rightsquigarrow x$ (and thus $x \leftrightsquigarrow x$).
Given  $x \in V$, we denote by $Out_{\overrightarrow{E}}(x)$ (or $Out(x)$ for short) the set $\{t \in V : t\neq x \land x \rightsquigarrow t\}$
and by $In_{\overrightarrow{E}}(x)$ (or $In(x)$ for short) the set $\{t \in V : t\neq x \land t\rightsquigarrow x\}$.

Given a dissimilarity space $(X, d)$ and  a tree $T=(X,E)$, an orientation $\overrightarrow{E}$ of $T$ is {\em compatible} for $d$ if, for every oriented path $p_1\rightarrow p_2\rightarrow\ldots\rightarrow p_k$, the space $(\{p_1, p_2,\ldots, p_k\}, d)$ is one-way-Robinson and admits $p_1<p_2<\ldots < p_k$ as a compatible order.  
We denote the number of oriented paths in a compatible orientation $\overrightarrow{E}$ as $\xi(\overrightarrow{E})$, where a path can include another one.
%The number of oriented paths in a compatible orientation $\overrightarrow{E}$ is denoted by $\xi(\overrightarrow{E})$ (a path can be included in another one). 
If $\overrightarrow{E}$ is a compatible orientation, we say that the oriented tree $\overrightarrow{T} = (X, \overrightarrow{E})$ is a {\em compatible oriented tree}.

An {\em optimal orientation} is  a compatible orientation $\overrightarrow{E}$ which maximizes $\xi(\overrightarrow{E})$. We will call $\xi(T, d)$ the number of oriented paths in an optimal orientation of $T$ for $d$. 

\section{An efficient algorithm to recognize two-way-Robinson dissimilarities}\label{sec:two-way-rob}
In this section, we present an efficient algorithm for recognizing two-way-Robinson spaces. Given a dissimilarity space $(X,d)$ and $x,y\in X$, we define the {\em segment} $S(x,y)$ as the set $\{t\in X : d(x,y) \geq \max\{d(x,t), d(t,y)\} \land d(y,x) \geq \max\{d(y,t) d(t,x)\}\} $

\begin{proposition}
    Let $(X,d)$ be a dissimilarity space. Then
    $(X,d)$ is two-way-Robinson
    if and only if
    there exists a total order $<$ on X such that, when $X$ is sorted along $<$, all segments $S(x,y)$ are intervals (for $<$).

    In this case, the sets $S(x,y)$ are intervals for all compatible orders of $(X, d)$.
\end{proposition}
\begin{proof}
    Suppose that $(X,d)$ is two-way-Robinson and $X$ is ordered along a compatible order.
    Let $x,y,t, u \in X$. With no loss of generality, we suppose that $x<y$. 
    If $x<t<y$, we have   $t \in S(x,y)$. 
    If $y<t < u$, then $d(x,u) \leq d(x,t))$ and $d(y,t) \leq d(y,t)$. If $u \in S(x,y)$, we have $d(x,t) \leq d(x,y)$ and $d(y,t) \leq d(y,x)$. So $t \in S(x,y)$.
    The case $u< t < x$ is similar, and so the segment $S(x,y)$ is an interval (for $<$).

    Conversely, suppose that $X$ is sorted along an order $<$ such that all segments $S(x,y)$ are intervals for $<$.
    As $x,y \in S(x,y)$, $\{t: x<t<y\} \subset S(x,y)$, {\it i.e.} $(X,d)$ is two-way-Robinson and $<$ is a compatible order.
\end{proof}

A $n\times m$ 01-matrix has the Consecutive Ones Property~\cite{booth1976testing} if its rows can be permuted in such a way that, in all columns, the 1's appear consecutively.
The algorithm of Booth and Lueker~\cite{booth1976testing} determines in $O(n\cdot m)$ if an $n\times m$ matrix has the Consecutive Ones Property and returns a PQ-tree which represents the permutations that make the ones consecutive.

Given a dissimilarity space $(X, d)$ with $\vert X\vert = n$, we can build, in $O(n^3)$, the $n\times(n^2-n)$ matrix $M$ with rows indexed by $X$ and columns by the segments $S(x,y)$ such that $M[i,j] = 1$ if the $i$th element  is in the $j$th segment and $M[i,j] = 0$ otherwise.
The algorithm of Booth and Lueker determines in $O(n^3)$ if $M$ has the Consecutive Ones Property or not. So we have:

\begin{theorem}\label{thm:two-way-rob}
    In time $O(\vert X\vert^3)$, we can determine if a dissimilarity space $(X, d)$ is two-way-Robinson.
\end{theorem}

Notice that a PQ-tree can represent the set of orders compatible with a (symmetric) Robinson dissimilarity. So, given a two-way-Robinson dissimilarity, a (symmetric) Robinson dissimilarity exists with the same compatible orders.

\section{NP-Hardness of the assignment and orientation problems}\label{sec:nphard-Or-As}

This section introduces two reductions establishing the NP-Completeness of the Assignment and Orientation problems. These reductions provide critical insights into the computational boundaries of the problems and open the door to efficient algorithmic approaches in some instances.

 As decision problems, we can settle the assignment and the orientation problems in the following way:
\begin{description}
 \item[]{\sc Assignment}:
 Given a dissimilarity space $(X, d)$ and an oriented tree $\overrightarrow{T} =(V, \overrightarrow{E})$ with $\vert X \vert = \vert V\vert$, does it exist a one-to-one map from $X$ to $V$ such that $\overrightarrow{E}$ is compatible\linebreak[2] for $d$?

\item[]{\sc Orientation}: 
Given a dissimilarity space $(V, d)$, a tree $T =(V, E)$ and an integer $\kappa$, does it exist a orientation $\overrightarrow{E}$, compatible for $d$, with $\xi(\overrightarrow{E}) \geq \kappa$?
\end{description}

\begin{theorem} \label{THEO_Orientation_NP}
    {\sc Orientation} is NP-complete.
\end{theorem}

\begin{proof}
Clearly, {\sc Orientation} is NP. 
We prove it is NP-hard with a reduction from {\sc 3-Sat}. 

Given a system $\mathcal{C}$ of $m$ clauses $c_1, c_2,\ldots, c_m$ (all with three literals) on $n$ variables \linebreak[4] $v_1,v_2,\ldots, v_n$, we consider the tree $\T_{SAT} = (V, E)$ where: 
\[
\begin{array}{rcl}
    V & = & \{y\} \cup \{x_i, 1\leq i\leq n\} \\
      & \cup & \{x^{i,+}_k,  1\leq i \leq n, 1\leq k\leq 7m+2\}\\
      & \cup &\{x^{i,-}_k,  1\leq i \leq n, 1\leq k\leq 7m+2\}\\
      & \cup &\{z^j_k, 1\leq j\leq m, 1\leq k\leq 7\}. 
\end{array}
\]
\[
\begin{array}{rcl}
E & = &  \{yx_i, 1\leq i\leq n\}\\
  & \cup & \{x_ix^{i,+}_k, 1\leq i\leq n, 1\leq k\leq 7m+2\}\\
  & \cup & \{x_ix^{i,-}_k, 1\leq i\leq n, 1\leq k\leq 7m+2\}\\
  &\cup & \{yz^j_k, 1\leq j \leq m, 1\leq k\leq 7\}.
\end{array}
\]
Figure~\ref{FIG_graph_SAT} shows such a tree. 
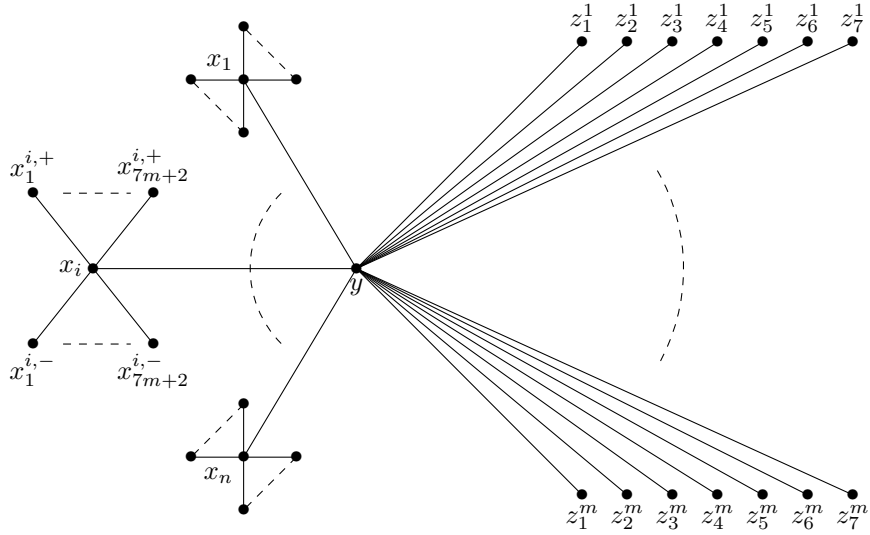
\begin{figure}[t]
\begin{center}
\begin{tikzpicture}
    \draw(0,0) node{\small{$\bullet$}} [below] node {\small{$y$}} ;
    \begin{scope}[xshift=-.5cm,yshift=1.5cm]
        \draw(-1, 1) node{\small{$\bullet$}} [above left] node {\small{$x_1$}} ;
        \draw(-1,.3) node{\small{$\bullet$}} ; \draw(-1,1) -- (-1,.3) ;
        \draw(-1.7,1) node{\small{$\bullet$}} ; \draw(-1,1) -- (-1.7,1) ;
        \draw(-.3, 1) node{\small{$\bullet$}} ; \draw(-1,1) -- (-.3,1) ;
        \draw(-1, 1.7) node{\small{$\bullet$}} ; \draw(-1,1) -- (-1,1.7) ;
        \draw[dashed] (-1, .3) -- (-1.7,1) ; \draw[dashed] (-.3,1) -- (-1,1.7) ;
    \end{scope}
    \begin{scope}[xshift=-.5cm,yshift=-1.5cm]
        \draw(-1, -1) node{\small{$\bullet$}} [below left] node {\small{$x_n$}} ;
        \draw(-1,-.3) node{\small{$\bullet$}} ; \draw(-1,-1) -- (-1,-.3) ;
        \draw(-1.7,-1) node{\small{$\bullet$}} ; \draw(-1,-1) -- (-1.7,-1) ;
        \draw(-.3, -1) node{\small{$\bullet$}} ; \draw(-1,-1) -- (-.3,-1) ;
        \draw(-1, -1.7) node{\small{$\bullet$}} ; \draw(-1,-1) -- (-1,-1.7) ;
        \draw[dashed] (-1, -.3) -- (-1.7,-1) ; \draw[dashed] (-.3,-1) -- (-1,-1.7) ;
    \end{scope}
    \draw (-1.5,2.5) -- (0,0) -- (-1.5,-2.5) ;
    \draw [dashed] (-1,1) arc (135:225:1.41) ;
    %\draw(-1,1) --(-1,-1) ;
    \begin{scope}
        \draw(-3.5, 0) node{\small{$\bullet$}} [left] node {\small{$x_i$}} ;
        \draw(-3.5,0) --(0,0) ;
        \draw(-3.5,0)--(-4.3,-1); \draw(-4.3,-1)node{\small{$\bullet$}} [below] node {\small{$x^{i,-}_1$}} ;
        \draw(-3.5,0)--(-2.7,-1); \draw(-2.7,-1)node{\small{$\bullet$}} [below] node {\small{$x^{i,-}_{7m+2}$}} ;
        \draw[dashed] (-3.9,-1) -- (-3,-1) ;
        \draw(-3.5,0)--(-4.3,1); \draw(-4.3,1)node{\small{$\bullet$}} [above] node {\small{$x^{i,+}_1$}} ;
        \draw(-3.5,0)--(-2.7,1); \draw(-2.7,1)node{\small{$\bullet$}} [above] node {\small{$x^{i,+}_{7m+2}$}} ;
        \draw[dashed] (-3.9,1) -- (-3,1) ;
    \end{scope}
    \draw(3,3) node{\small{$\bullet$}} [above] node {\small{$z^1_1$}} ; \draw(0,0)--(3,3) ;
    \draw(3.6,3) node{\small{$\bullet$}} [above] node {\small{$z^1_2$}} ;  \draw(0,0)--(3.6,3) ;
    \draw(4.2,3) node{\small{$\bullet$}} [above] node {\small{$z^1_3$}} ; \draw(0,0)--(4.2,3) ;
    \draw(4.8,3) node{\small{$\bullet$}} [above] node {\small{$z^1_4$}} ;  \draw(0,0)--(4.8,3) ;
    \draw(5.4,3) node{\small{$\bullet$}} [above] node {\small{$z^1_5$}} ; \draw(0,0)--(5.4,3) ;
    \draw(6,3) node{\small{$\bullet$}} [above] node {\small{$z^1_6$}} ; \draw(0,0)--(6,3) ;
    \draw(6.6,3) node{\small{$\bullet$}} [above] node {\small{$z^1_7$}} ; \draw(0,0)--(6.6,3) ;
    \draw(3,-3) node{\small{$\bullet$}} [below] node {\small{$z^m_1$}} ; \draw(0,0)--(3,-3) ;
    \draw(3.6,-3) node{\small{$\bullet$}} [below] node {\small{$z^m_2$}} ; \draw(0,0)--(3.6,-3) ;
    \draw(4.2,-3) node{\small{$\bullet$}} [below] node {\small{$z^m_3$}} ; \draw(0,0)--(4.2,-3) ;
    \draw(4.8,-3) node{\small{$\bullet$}} [below] node {\small{$z^m_4$}} ; \draw(0,0)--(4.8,-3) ;
    \draw(5.4,-3) node{\small{$\bullet$}} [below] node {\small{$z^m_5$}} ; \draw(0,0)--(5.4,-3) ;
    \draw(6,-3) node{\small{$\bullet$}} [below] node {\small{$z^m_6$}} ; \draw(0,0)--(6,-3) ;
    \draw(6.6,-3) node{\small{$\bullet$}} [below] node {\small{$z^m_7$}} ; \draw(0,0)--(6.6,-3) ;
    \draw [dashed] (4,1.3) arc(30: -30 : 2.6) ;
\end{tikzpicture}
\caption{The tree $\T_{SAT}$ for a 3-cnf formula with  $m$ clauses $c_1, c_2,\ldots, c_m$, all with three literals, on $n$ variables $v_1,v_2,\ldots, v_n$. \label{FIG_graph_SAT}}
\end{center}
\end{figure}

We now construct a dissimilarity $d_{SAT}$ on $V$. For $x\neq y$, $d_{SAT}(x,y) \in \{1,2\}$. If $xy$ is an edge of $\T_{SAT}$, $d_{SAT}(x,y)=2$. So, if $d_{SAT}(x,y) = 1$, then, for any orientation $\overrightarrow{E}$ of $E$, we have $x\not\leftrightsquigarrow y$, in particular, if there exists $z$ with $xz, yz \in E$, we may, in $\overrightarrow{E}$, have $x\leftarrow z \rightarrow y$ or $x\rightarrow z \leftarrow y$, but neither $x\rightarrow z\rightarrow y$ nor $y \rightarrow z\rightarrow x$.
We now give the pairs $xy \in V$ for which $d_{SAT}(x,y) = 1$ (the value of $d_{SAT}$ will be $2$ for all the other pairs).
\begin{itemize}
    \item $\forall i,i' \in\{1,\dots,n\}$, $d_{SAT}(x_i,x_{i'}) = 1$. So all the edges $x_iy$ are oriented similarly (towards $y$ or from $y$). With no loss of generality, we will suppose that for all $i$, we have $x_i\rightarrow y$.
    \item $\forall i \in \{1,\ldots, n\}, k,k' \in \{1,\ldots,7m+2\}$, \[d_{SAT}(x^{i,+}_k, x^{i,+}_{k'}) = d_{SAT}(x^{i,-}_k, x^{i,-}_{k'}) = 1.\]
    As above, all edges $x^{i,+}_kx_i$ are oriented in the same way (towards or from $x_i$), and we have the same property for the edges $x^{i,-}_kx_i$. So, for all $i\in \{1,\ldots,n\}$, we are in one of the following cases:
    \begin{enumerate}
        \item $\forall k,k'\in\{1,\ldots,2m+2\}$, $x^{i,+}_k \rightarrow x_i \rightarrow x^{i,-}_{k'}$
        \item $\forall k,k'\in\{1,\ldots,2m+2\}$, $x^{i,-}_k \rightarrow x_i \rightarrow x^{i,+}_{k'}$
        \item $\forall k,k'\in\{1,\ldots,2m+2\}$, $x^{i,+}_k \rightarrow x_i \leftarrow x^{i,-}_{k'}$
        \item $\forall k,k'\in\{1,\ldots,2m+2\}$, $x^{i,+}_k \leftarrow x_i \rightarrow x^{i,-}_{k'}$
    \end{enumerate}
    We now count the number of paths of length $>1$ (the number of paths of length $1$ is the same for any orientation) involving the vertices $x^{i,+}_k$ or $x^{i,-}_k$ in these different cases. 
    \begin{itemize}
        \item In Case 4, the vertices $x^{i,+}_k$ and $x^{i,-}_k$ are in no paths of length $>1$
    
        \item In cases 1 and 2,  $x^{i,+}_k$ and $x^{i,-}_k$ are involved in $(7m+2)^2 + (7m+2)\cdot (K+1)$ paths, where $K$ is the number of vertices which are reachable from $y$. The $(7m+2)^2$ paths are the paths between the $x^{i,+}_k$ and the $x^{i,-}_k$ and the $(7m+2)\cdot (K+1)$ are the paths from (in Case 1) the $x^{i,+}_k$'s to $y$ and the $K$ $z_k^j$'s which are reachable from $y$.
    
        \item In Case 3, there are $2\cdot(7m+2)\cdot(K+1)$ paths from the $x^{i,+}_k$'s and $x^{i,-}_k$'s to the other vertices.
    \end{itemize}
    Our goal is to get more than $\kappa$ paths. So, by taking $\kappa$ great enough (we will determine its value later), we can force, since $K\leq 7m$, that we are in Case 1 or 2 for all $i$. Case 1 will correspond to $v_i =$ {\bf True} and Case 2 to $v_i=$ {\bf False}.
    
    \item There are seven ways for a clause $c_j$ with 3 literals to be satisfied: the first literal is the only one which is true, the second literal is the only one which is true,\ldots, the first and the third literals are true and the second literal is false,\ldots, the three literals are true. The seven vertices $z^j_1,\ldots, s^j_7$ will represent these seven cases. More precisely, if, for example, $c_j = v_{i} \lor \neg v_{i'} \lor v_{i''}$ (the other cases are similar), we set, for all $1\leq k\leq 7m+2$:
    \begin{itemize}
        \item $d_{SAT}(z^j_1, x^{i,-}_k) = d_{SAT}(z^j_1, x^{i',-}_k) = d_{SAT}(z^j_1, x^{i'',+}_k) = 1$ 
        
        \hspace{1cm}$z^j_1$ represents the case  {\it the first literal is the only true}.

        \item $d_{SAT}(z^j_2, x^{i,+}_k) = d_{SAT}(z^j_2, x^{i',+}_k) = d_{SAT}(z^j_2, x^{i'',+}_k) = 1$ 

        \hspace{1cm}$z^j_2$ represents the case  {\it the second literal is the only one true}.

        \item $d_{SAT}(z^j_3, x^{i,+}_k) = d_{SAT}(z^j_3, x^{i',-}_k) = d_{SAT}(z^j_3, x^{i'',-}_k) = 1$ 

        \hspace{1cm}$z^j_3$ represents the case  {\it the third literal is the only true}.

        \item $d_{SAT}(z^j_4, x^{i,-}_k) = d_{SAT}(z^j_4, x^{i',+}_k) = d_{SAT}(z^j_4, x^{i'',+}_k) = 1$ 

        \hspace{1cm}$z^j_4$ represents the case  {\it the first and the second literals are the only true ones}.

        \item $d_{SAT}(z^j_5, x^{i,-}_k) = d_{SAT}(z^j_5, x^{i',-}_k) = d_{SAT}(z^j_5, x^{i'',-}_k) = 1$ 

        \hspace{1cm}$z^j_5$ represents the case  {\it the first and the third literals are the only true ones}.

        \item $d_{SAT}(z^j_6, x^{i,+}_k) = d_{SAT}(z^j_6, x^{i',+}_k) = d_{SAT}(z^j_6, x^{i'',-}_k) = 1$ 

        \hspace{1cm}$z^j_6$ represents the case  {\it the second and the third literals are the only true ones}.

        \item $d_{SAT}(z^j_7, x^{i,-}_k) = d_{SAT}(z^j_7, x^{i',+}_k) = d_{SAT}(z^j_7, x^{i'',-}_k) = 1$ 

        \hspace{1cm}$z^j_4$ represents the case  {\it the  literals are all true}.
    \end{itemize}
    In any compatible orientation of $\T$ and for all $1\leq j\leq m$, there is at most one edge $yz^j_\ell$ which is oriented as $y\rightarrow z^j_\ell$: suppose, by way of contradiction, that $y\rightarrow z^j_\ell$ and $y \rightarrow z^j_{\ell'}$. Since there exists $i\in\{1,\ldots,n\}$ with, for all $1\leq k\leq 7m+2$, $d_{SAT}(x^{i,+}_k, z^j_\ell) = d_{SAT}(x^{i,-}_k, z^j_{\ell'}) = 1$, we have, for all $k\in\{1,\ldots, 7m+2\}$, $x_i\rightarrow x^{i,-}_k$ and $x_i\rightarrow x^{i,+}_k$. Such an orientation will not satisfy $\xi(\overrightarrow{E}) \geq \kappa$ for $\kappa$ great enough.
\end{itemize}
We now set
\[
\begin{array}{rclr}
    \kappa & := & 7m + 5n + 14nm \,\,\, & \text{ the number of edges} \\
           &  +  &    n\cdot(7m+2)^2  & \text{ for the paths } x^{i,\pm}_k\rightarrow x_i \rightarrow x^{i,\mp}_k \\
           &  + &    nm & \text{ for the paths } x_i \rightarrow y\rightarrow z^j_\ell \\
           & + & n\cdot(7m+2) & \text{ for the paths } x^{i,\pm}_k \rightsquigarrow y\\
           & + & nm\cdot(7m+2) & \text{ for the paths } x^{i,\pm}_k \rightsquigarrow z^j_\ell \\
           & + & 6m^2 & \text{ for the paths } z^j_\ell \rightarrow y \rightarrow z^{j'}_{\ell'}
\end{array}   
\]
An affection $\mathcal{A}$ to {\bf True} or {\bf False} of all the variables $v_i$ corresponds to an orientation of all edges $x_ix^{i,+}_k$ and $x_ix^{i,-}_k$ following Case 1 or 2, and a clause $v_j$ is satisfied by $\mathcal{A}$ if and only if there exists a vertex $z^j_\ell$ such that we can orient the edge $yz^j_\ell$ as $z\rightarrow z^j_\ell$. So $\mathcal{C}$ is satisfiable if and only if $\xi(\T_{SAT}, d_{SAT}) = \kappa$.
\end{proof}

Notice that, in the proof of Theorem~\ref{THEO_Orientation_NP}, we always construct (up to an isomorphism) the same oriented tree.
It is thus easy to adapt this proof to prove the NP-completeness of the {\sc Assignment} problem.
Instead of that, we will prove the stronger Corollary~\ref{CORO_Assignment_NP} (which is false for the {\sc Orientation} problem -- see Proposition~\ref{PROP_algo_orientation_path}). 
 Before proving this, we have to consider another problem:
\begin{description}
    \item[] {\sc Robinson-Subset}: Given a dissimilarity space $(X, d)$ and an integer $\kappa$, does it exist  $X'\subset X$ with $\vert X'\vert = \kappa$ such that $(X',d)$ is Robinson?
\end{description}

\begin{lemma}
    {\sc Robinson-Subset} is NP-complete.
\end{lemma}
\begin{proof}
    {\sc Robinson-Subset} is clearly NP. We prove that it is NP-hard by reduction from Hamiltonian Path.
    Let $G=(V,E)$ be a graph with $ V = \{v_1,\ldots, v_n\}$ and $E = \{e_1,\ldots, e_m\}$. We consider the set $X := \{x_i^k, 1\leq i\leq n, 1\leq k\leq m+1\} \cup \{y_j, 1\leq j \leq m\}$ and the dissimilarity $d$ on $X$ defined by:
\begin{itemize}
\item[-]     $\forall i\in\{1,\ldots, n\}, k,k' \in \{1,\ldots, m+1\}$, $d(x_i^k, x_i^{k'}) = 1$
    
 \item[-]    If $e_j = v_iv_{i'}$, then $\forall 1\leq k\leq m+1$, $d(y_j, x_i^k) = d(y_j, x_{i'}^k) = 1$

 \item[-]    All other values of $d$ are 2.
    \end{itemize}
    If $G$ is Hamiltonian (we suppose, with no loss of generality, that the Hamiltonian path is $v_1,v_2,\ldots,v_n$ and that the involved edges are, in this order, $e_1, e_2,\ldots, e_{n-1}$), then the set $\{x_i^k, 1\leq i\leq n, 1\leq k\leq m+1\} \cup \{y_j, 1\leq j \leq n-1\}$ is Robinson (it admits $x_1^1<\ldots x_1^{m+1}<y_1<x_2^1<\ldots <x_2^{m+1}< y_2<\ldots <y_{n-1}<x_n^1<\ldots x_n^{m+1}$ as a compatible order) and of size $n\cdot(m+1) + n-1$.

    Conversely, suppose that $X$ has a  subset $X'$ with $\vert X'\vert \geq n\cdot(m+1) + n-1$ such that $(X',d)$ is Robinson. We denote by $X'_i$ the set $\{x_i^k, 1\leq k\leq m+1\} \cap X'$. We have:
\begin{itemize}
    \item[] \textbf{Claim 1.} $\forall 1\leq i\leq n$, $ X'_i \neq \emptyset$.

    Otherwise, there is less than $n\cdot(m+1) + n-1$ points in $X'$.

    \item[] \textbf{Claim 2.} {\it If the edge $e_j$ is incident with $v_i$ and $y_j\in X'$, then, for any compatible order of $X'$, $X'_i \cup\{y_j\}$ is an interval.}

    Suppose that, for a compatible order $\sigma$, there is a point $z \in X'\setminus  X'_i$ between $y_j$ and a point $x_i^k\in X'_i$. If $z = y_{j'}$, then $d(y_j, z) = 2$, otherwise $d(x_i^k, z) = 2$. Both are impossible since $d(x_i^k, y_j) = 1$.

    \item[] \textbf{Claim 3.} {\it If $e_i, e_j, e_k$ are incident to a same vertex $v_\ell$, then $\{y_i, y_j, y_k\} \not\subset X'$}.

    If $\{y_i, y_j, y_k\} \subset X'$, then, for every compatible order 
    $\sigma$, $X'_\ell\cup \{y_i\}$, $X'_\ell\cup \{y_j\}$ and $X'_\ell\cup \{y_k\}$ are intervals, a contradiction.

    \item[] \textbf{Claim 4.} {\it If $v_{i_1},v_{i_2},\ldots, v_{i_k}$ is a cycle with $e_{i_k}=v_{i_k}v_{i_1}$ and, for all $1\leq j< k$, $e_{i_j} = v_{i_j}v_{i_{j+1}}$,  then $\{y_{i_1}, y_{i_2},\ldots, y_{i_k}\} \not\subset X'$}.

    If $\{y_{i_1}, y_{i_2},\ldots, y_{i_k}\} \subset X'$, then $X'_{i_1} \cup \{y_{i_k}, y_{i_1}\}$ and  the sets $X'_{i_\ell} \cup \{y_{i_\ell}, y_{i_{\ell-1}}\}$, for $2\leq \ell \leq k$ are all intervals, a contradiction.
\end{itemize}
By Claims 3 and 4, $X'$ corresponds to a collection of paths (there are neither cycles nor vertices of degree $>2$). The only solution for $\vert X'\vert$ to be $\geq n\cdot(m+1) + n-1$ is that, for all $1\leq i\leq n$, $\vert X'_i\vert = m+1$ and that there is only one path. This path is Hamiltonian.
\end{proof}

\begin{theorem} \label{CORO_Assignment_NP}
    {\sc Assignment} is NP-complete, even when restricted on paths.
\end{theorem}

\begin{proof}
    Clearly, {\sc Assignment} is NP. We prove it is NP-hard by a reduction from the problem {\sc Robinson-Subset}. Given an instance $((X,d), \kappa)$ of {\sc Robinson-Subset},
    %where $X=\{x_1,\ldots, x_n\}$, 
    we consider the path $P = v_1-v_2-\ldots-v_n$, oriented as $\overrightarrow{P} = v_1\rightarrow v_2\rightarrow \ldots\rightarrow v_\kappa \leftarrow v_{\kappa+1} \rightarrow v_{\kappa+2}\ldots v_n$.
    There is a one-to-one map between $X$ and $\{v_1,\ldots, v_n\}$ such that $\overrightarrow{P}$ is a compatible orientation for $d$ if and only if $X$ has a Robinson subset of size $\kappa$.
\end{proof}

\section{Optimal orientation when all paths are Robinson}\label{sec:optimalAllPathRob}
%in some instances}

In this section, we focus on the scenario where, for a given dissimilarity space $(X, d)$ and a tree $T = (X, E)$, every path within $T$ is Robinson. This condition is satisfied when $(X, d)$ is a T-Robinson space and $T$ represents one of its corresponding trees. A particular case of this situation is when $d$ is a constant dissimilarity on $X$. Under these constraints, we propose an $O(n^2)$ algorithm to solve the Orientation problem efficiently. In this case, an optimal orientation is an orientation of $T$, which maximizes the number of pairs $\{x, y\}$ such that  $x \leftrightsquigarrow y$.

We first give (\Cref{SUB_prop_opt_orientation}) some properties of an optimal orientation, from which we derive (\Cref{SUB_opt_algo}) an optimal algorithm. %Finally (\Cref{SUB_bounding}), we give the minimum (and maximum) number of orientations that a tree with $n$ vertices can have.

\subsection{Properties of an optimal orientation} \label{SUB_prop_opt_orientation}

Given an orientation $\overrightarrow{E}$ of a tree $T=(V, E)$, a vertex $x$ is {\em central} (for $\overrightarrow{E}$) if, for every vertex $y$, we have $x \leftrightsquigarrow y$.

\begin{lemma} \label{LEM_central_orientation}
    An orientation $\overrightarrow{E}$ of a tree $T=(V, E)$ has a central vertex if and only if:
     \begin{equation} \label{EQU_def_central}
     \forall\, t,t' \in V, \exists\, c_{tt'} \in V : t \leftrightsquigarrow c_{tt'} \land t' \leftrightsquigarrow c_{tt'}
     \end{equation}
 \end{lemma}

\begin{proof}
    We prove the ``if" part by induction on $n$. Suppose that the property is true for all trees with $n$ vertices and let $T= (V, E)$ be a graph with $n+1$ vertices with an orientation $\overrightarrow{E}$  such that \cref{EQU_def_central} holds.
    Let $u$ be a leaf of $T$ and $x$ its neighbor. With no loss of generality, we suppose that $x\rightarrow u$ (otherwise, we reverse the orientation of all edges of $E$). Let $T'$ be the subgraph induced by $V':=V\setminus \{u\}$ and $\overrightarrow{E'}$ the orientation $\overrightarrow{E}$ restricted on $T'$.
    
    \cref{EQU_def_central} holds for $\overrightarrow{E'}$: if, for $t, t' \in V'$, $c_{t,t'} = u$, we can set $c_{t,t'} := x$. So, by the induction hypothesis, $\overrightarrow{E'}$ has a central vertex $c$. If   $c \rightsquigarrow x$ in $\overrightarrow{E'}$, then $c\rightsquigarrow u$ in $\overrightarrow{E}$ and $c$ is a central vertex in $\overrightarrow{E}$.
    Otherwise:
    \begin{itemize}
    \item If for all $t\in V\setminus\{u,x\}$, $t\leftrightsquigarrow x$, then $x$ is a central vertex of $\overrightarrow{E}$,
    
    \item If there exists $t$ such that $t \not\leftrightsquigarrow x$, then \cref{EQU_def_central} is not verified for the pair $\{t, u \}$ in $\overrightarrow{E}$, a contradiction.
    \end{itemize}
    The ``only if" part is obvious by taking, as $c_{tt'}$, the central vertex.
\end{proof}

\begin{lemma} \label{LEM_optimal_central}
    An optimal orientation  $\overrightarrow{E}$ of a tree $T$ has a central vertex.
\end{lemma}
\begin{proof}
    By contradiction, suppose that $\overrightarrow{E}$ is optimal but has no central vertex. So, by \cref{LEM_central_orientation}, there exist two vertices $u$ and $v$ such that, for all vertex $t$, if there exists a directed path between $t$ and $u$, there is no directed path between $t$ and $v$ and reciprocally (there may exist a vertex $t$ with no directed path between $u$ and $t$ and between $v$ and $t$). We take $u$ and $v$ the closest possible.
    In the non-directed path $P$ between $u$ and $v$, let $x$ be the neighbor of $u$ and $y$ the neighbor of $v$. 
    Notice that $x\neq y$ (otherwise, we would have $x \leftrightsquigarrow u$ and $x \leftrightsquigarrow v$).
    We suppose, with no loss of generality, that $x\rightarrow u$.
    We set $P = (u, x, p_1, \ldots, p_k, y, v)$.

\begin{itemize}
    \item[] \textbf{Claim.} 
    {\it In $\overrightarrow{E}$, we have $x\rightsquigarrow y$ and $v \rightarrow y$.}
        
        First, notice that if $y \rightsquigarrow x$, then $y$ is reachable from $v$ and can reach $u$, which contradicts the hypothesis.
        Suppose that $p_1\rightarrow x$. Let $i$ be the greatest index such that $p_i\rightsquigarrow x$.
        If a vertex $t$ can reach $p_{i-1}$, it can reach $u$, so there is no directed path between $t$ and $v$. If $t$ can be reached from $p_{i-1}$, there is no directed path between $t$ and $y$ ($p_i$ is an impassable border). Since the distance between $u$ and $v$ is minimal, we would have taken $p_{i-1}$ instead of $u$. 

        So $x\rightarrow p_1$. Let $j$ be the greatest index such that $x \rightsquigarrow p_j$. If $p_j\neq y$, then, as above, we would have taken $p_{j+1}$ instead of $v$.
        So there exists a path $x \rightsquigarrow y$.

        If $y\rightarrow v$, $x$ would be able to reach both $u$ and $v$, a contradiction with the hypothesis.
\end{itemize}
    We get the situation of \cref{FIG_two_centrals_bis}, where:
    $\mathcal{O}(x)$ is the set of vertices reachable from $x$ but not from another vertex of $P$ ($u \in \mathcal{O}(x)$) and $\mathcal{I}(y)$ is the set of vertices that can reach $y$ and no other vertex in $P$ ($v \in \mathcal{I}(y)$). Recall that
    $In(x)$ is the set of vertices that can reach $x$ and that
    $Out(y)$ is the set of vertices that can be reached from $y$. 
%    $In(P)$ is the set of points that can reach $p_k$ but not $x$.
    Notice that, for   $1\leq i\leq k$, there is no vertex $t$  with $p_i \rightarrow t$ (we would have taken $t$ instead of $u$) nor $t\rightarrow p_i$ (we would have taken $t$ instead of $u$).
    
    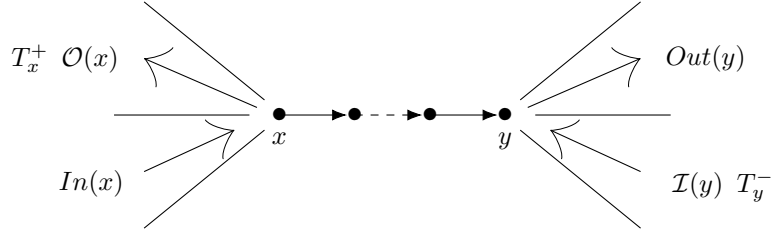
\begin{figure}[t]
    \begin{center}
    \begin{tikzpicture}    
        \draw (0,0) node{\large{$\bullet$}} [below=1mm] node{\small{$x$}} ;
        \draw [-{Latex[length=2mm]}] (0,0) -- (.9,0) ;
        \draw (1,0) node{\large{$\bullet$}} ;
        \draw [dashed, -{Latex[length=2mm]}] (1,0) -- (1.9,0) ;
        \draw (2,0) node{\large{$\bullet$}} ;
        \draw [-{Latex[length=2mm]}] (2,0) -- (2.9,0) ;
        \draw (3,0) node{\large{$\bullet$}} [below=1mm] node{\small{$y$}} ;
        %\draw(1, .5) -- (.5, 2) ;
        %\draw(2, .5) -- (2.5, 2) ;
        %\draw [-{Classical TikZ Rightarrow[length=3mm]}](1.5, .4) -- (1.5, 1.4) [above = 2mm] node{\small{$Out(P)$}} ;
        %\draw(1, -.5) -- (.5, -2) ;
        %\draw(2, -.5) -- (2.5, -2) ;
        % \draw [-{Classical TikZ Rightarrow[length=3mm]}](1.5, -1.4) -- (1.5, -.4) [below = 12mm] node{\small{$In(P)$}} ;
         \draw(-2.2, 0) -- (-.4, 0) ;
         \draw [-{Classical TikZ Rightarrow[length=3mm]}](-.3, .1) -- (-1.8, .75) [left = 2mm] node{\small{$\T_x^+ \,\,\,\mathcal{O}(x)$}} ;
         \draw (-1.8, 1.5) -- (-.2, .2) ;
         \draw [-{Classical TikZ Rightarrow[length=3mm]}](-1.8, -.75) -- (-.6, -.2) [left = 19mm, below = 4mm] node{\small{$In(x)$}} ;
         \draw (-1.8, -1.5) -- (-.2, -.2) ;
         
         \draw [-{Classical TikZ Rightarrow[length=3mm]}](3.3, .1) -- (4.8, .75) [right = 2mm] node{\small{$Out(y)$}} ;

         \draw [-{Classical TikZ Rightarrow[length=3mm]}](4.8, -.75) -- (3.6, -.2) [right = 23mm, below = 4mm] node{\small{$\mathcal{I}(y) \,\,\, \T_y^-$}} ;

          \draw(5.2, 0) -- (3.4, 0) ;
         %\draw [-{Classical TikZ Rightarrow[length=3mm]}](-.3, .1) -- (-1.8, .75) [left = 2mm] node{\small{$Out(P)$}} ;
         \draw (4.8, 1.5) -- (3.2, .2) ;
         %\draw [-{Classical TikZ Rightarrow[length=3mm]}](-1.8, -.75) -- (-.6, -.2) [left = 19mm, below = 4mm] node{\small{$In(P)$}} ;
         \draw (4.8, -1.5) -- (3.2, -.2) ;
    \end{tikzpicture}
    \end{center}
    \caption{The neighborhood of  $x$ and $y$
        \label{FIG_two_centrals_bis}}
    \end{figure}

    Let $\T_x^+$ be the greatest subtree of $\T$ containing $x$ but no other vertex in $P$ and no vertices in $In(x)$ and $\T_y^-$ be the greatest subtree containing $y$ and no other vertices in $P$ and no vertices in $Out(y)$.
    By reversing the orientation of all edges in $\T_x^+$, we get from $\overrightarrow{E}$ a new orientation $\overrightarrow{E_x}$ and we have:
    \[\xi(\overrightarrow{E_x}) - \xi(\overrightarrow{E}) = \vert \mathcal{O}(x)\vert \cdot(\vert Out(y)\vert + k + 1 - \vert In(x)\vert)\]
    By reversing the orientation of all arcs in $\T_y^-$, we get a orientation $\overrightarrow{E_y}$ and we have:
    \[\xi(\overrightarrow{E_y}) - \xi(\overrightarrow{E}) = \vert \mathcal{I}(y)\vert\cdot (\vert In(x)\vert + k + 1 - \vert Out(y) \vert   )\]
    As $\vert \mathcal{O}(x)\vert, \vert \mathcal{I}(y)\vert > 0$ and $k\geq 0$, at least one of $\xi(\overrightarrow{E_x}) - \xi(\overrightarrow{E})$ and $\xi(\overrightarrow{E_y}) - \xi(\overrightarrow{E})$ is positive, a contradiction with the optimality of $\overrightarrow{E}$.
\end{proof}

\begin{proposition} \label{PROP_best_orientation_with_fixed_central}
    Let $\T=(V,E)$ be a tree and $x \in V$. An orientation $\overrightarrow{E}$ of $\T$ admitting $x$ as central vertex  maximizes $\xi$ if and only if it minimizes $\vert \vert Out(x)\vert - \vert In(x)\vert \vert$. 
    We denote by $\xi(x)$ this maximal value for $\xi$.
\end{proposition}
\begin{proof}
    If $x$ is a central vertex,  the number of paths of $\overrightarrow{E}$ inside a connected component of $\T\setminus x$ does not depend on $\overrightarrow{E}$. So the number of paths in $\overrightarrow{E}$ is, up to an additive constant, $In(x) \cdot Out(x)$.
    As  $In(x) = n-1-Out(x)$, this value is maximum when $\vert \vert Out(x)\vert - \vert In(x)\vert \vert$ is minimum.
\end{proof}

\begin{lemma} \label{LEM_best_central}
    Let $\T=(V,E)$ be a tree and $x\in V$. If a neighbor $y$ of $x$ is such that there are more than $n/2$ vertices closer from y than from $x$, then $\xi(y) \geq \xi(x) $.
\end{lemma}

\begin{proof}
    By \cref{PROP_best_orientation_with_fixed_central},   in one of the best orientations of $T$ with $x$ as central vertex, there is a path from $x$ to every vertex which is closer from $y$ than from $x$, and a path from all other vertices to $x$ (the only other best orientation of $T$ is the reverse of this one). This orientation also has $y$ as the central vertex. So $\xi(y)\geq \xi(x)$.
\end{proof}

\begin{proposition} \label{PROP_best_central}
    Let $\T=(V,E)$ be a tree with $\vert V\vert = n$ and $x\in V$. If, for each neighbor $y$ of $x$, there are at most $n/2$ vertices which are closer from $y$ than from $x$, then $\xi(T) = \xi(x)$.
\end{proposition}
\begin{proof}
    Suppose that $\T\setminus x$ is made of $p$ trees $\T_i = (V_i,E_i)$,   then, by \cref{PROP_best_central}, $\vert V_i\vert \leq n/2$ for each $i$ and so $\sum_{j\neq i}\vert V_j\vert \geq n/2-1$ (the $-1$ is due to $x$).
    Let $y\in V_i$ be a vertex which is central for an optimal orientation $\overrightarrow{\T}$ of $\T$ and let $x, z_1, \ldots, z_p=y$ be the path in $\T$ between $x$ and $y$.

    If $p>1$, there are more than $n/2$ vertices closer to $z_{p-1}$ than from $y$, a contradiction.

    If $p=1$, then $x$ is also a central vertex for $\overrightarrow{\T}$.
\end{proof}

\subsection{An optimal algorithm}
\label{SUB_opt_algo}

From \cref{PROP_best_orientation_with_fixed_central,PROP_best_central}, it is easy to derive an algorithm that finds an optimal orientation of a tree $\T$ for the constant dissimilarity:
\begin{enumerate}
    \item Find a vertex $x_c$ such that none of the subtrees we get by removing $x_c$ has more than $n/2$ vertices.
    \item Orient each of the connected components of $\T\setminus x_c$ (all edges in a component are oriented the same way: from $x_c$ or to $x_c$) such that $\vert \vert Out(x_c)\vert - \vert In(x_c)\vert \vert$ is minimal.
\end{enumerate}

\begin{proposition} \label{PROP_complexity_step1}
    Step 1 can be done in $O(n)$.
\end{proposition}
\begin{proof}
    First, we choose (at random) a vertex $x$. Then we compute, for each neighbor $y$ of $x$, $\theta_x(y)$ the number of vertices in the maximal subtree of $\T$ containing $y$ and not $x$: for any vertex $u$ and neighbor $v$ of $z$ have $\theta_u(v) = 1 + \sum_{t\in \Gamma_u(v)}\theta_v(t)$, where $\Gamma_u(v)$ is the set of the neighbors of $v$ different from $u$.
    So, if all the $\theta_x(y)$ are $<n/2$, $x_c = x$. Otherwise, we take as new candidate for $x_c$ the (unique) neighbor $y$ of $x$ such that $\theta_x(y) \geq n/2$.
    As to compute $\theta_x(y)$, we have calculated $\theta_y(t)$ for all $t\neq x$, we can determine if $x_c= y$ or if there is a neighbor $t$ of $y$ such that $\theta_y(t) \geq n/2$ (such a neighbor of $y$ can not be $x$), and so, by repeating this operation, we can find $x_c$ in $O(n)$.
\end{proof}

\begin{proposition} \label{PROP_complexity_step2}
    Step~2 can be done in $O(n\cdot \deg(x_c))$.
\end{proposition}
\begin{proof}
    We use a similar way to the classical pseudo-polynomial algorithm for the knapsack problem:
    we order (randomly) the neighbors of $x_c$ in $(y_1,\ldots, y_{\deg(x_c)})$ and we build a matrix $\mathcal{M}$ whose lines range from 0 to $n/2$ and  columns from 0 to $\deg(x_c)$ and such that $\mathcal{M}[i, j]$ is the greatest value of $|In(x_c)|$, which is $\leq i$, for $In(x_c) \subset \{y_1,\ldots, y_j\}$. This is done by Algorithm~\ref{ALGO_optimal_neigborhood_partition} \textsc{Optimal-Partition-of-Neighbors}.
  
    Each element of $\mathcal{M}$ is computed in $O(1)$, and so $\mathcal{M}$ is computed in $O(n\cdot\deg(x_c))$. From $\mathcal{M}$, we can determine the optimal partition of the neighbors of $x_c$ into $In(x_c), Out(x_c)$ in $O(deg(x_c)$.
\end{proof}

\begin{algorithm2e}
  \caption{\textsc{Optimal-Partition-of-Neighbors}}
  \label{ALGO_optimal_neigborhood_partition}
  \KwIn{A set $\{y_1, \ldots, y_p\}$ (the neighbors of $x_c$) and numbers $\theta(y_j)$ for $1 \leq j \leq p$.}
  \KwOut{A matrix $\mathcal{M}$ with $\lfloor n/2 \rfloor + 1$ rows and $p + 1$ columns.}
  \BlankLine
  \For{$0 \leq i \leq \lfloor n/2 \rfloor$}{
    $\mathcal{M}[i,0] \gets 0$\;
  }
  \For{$1 \leq j \leq p$}{
    \For{$0 \leq i \leq \lfloor n/2 \rfloor$}{
      \eIf{$\theta(y_j) > i$}{
        $\mathcal{M}[i,j] \gets \mathcal{M}[i, j-1]$\;
      }{
        $\mathcal{M}[i,j] \gets \max\{\mathcal{M}[i, j-1], \mathcal{M}[i-\theta(y_j), j-1] + \theta(y_j)\}$\;
      }
    }
  }
\end{algorithm2e}
\begin{comment}

\begin{algorithm}[htb]
  \caption{\textsc{Optimal-Partition-of-Neighbors}
    \label{ALGO_optimal_neigborhood_partition}}
    %{\small{
  \KwIn{
    A set $\{y_1, \ldots y_p\}$ (the neighbors of $x_c$) and numbers $\theta(y_j)$ for $1\leq j \leq p$.
  }
  \KwOut{
    A matrix $\mathcal{M}$ with $\lfloor n/2\rfloor +1$ lines and $p+1$ columns.
  }
  \Begin{
    \ForAll{$0 \leq i \leq \lfloor n/2\rfloor$}
        {$\mathcal{M}[i,0] \gets 0$ \;
        }
    \ForAll{$1\leq j \leq p$}
        {\ForAll{$0 \leq i \leq \lfloor n/2\rfloor$}
            {
            \eIf{$\theta(y_j) > i$}
                {$\mathcal{M}[i,j] \gets \mathcal{M}[i, j-1]$
                }
                {$\mathcal{M}[i,j] \gets \max\{\mathcal{M}[i, j-1], \mathcal{M}[i-\theta(y_j), j-1] + \theta(y_j)] \}$
                }
            }
        }
    }
\end{algorithm}
\end{comment}

From \cref{PROP_complexity_step1,PROP_complexity_step2}, we derive:

\begin{theorem}\label{Theo_optimal_orientation}
    Given a dissimilarity space $(X, d)$ with $\vert X\vert = n$ and a tree $\T$ with vertex set $X$, such that all paths in $\T$ are Robinson. 
    Then, it is possible to compute an optimal orientation of $\T$ in $O(n^2)$.
\end{theorem}

\section{Tractable instances for symmetric dissimilarities}\label{SECTION_tract_cases_non_constant}

In this section, we study two cases 
when $d$ is a symmetric but non-constant dissimilarity on a set $X$.
In \Cref{SUB_petals_stars}, we give,  when $\T=(X, E)$ is a star, an optimal $O(n^2)$ algorithm for the orientation problem, which can be turned into an efficient $O(n^3)$ algorithm for the assignation problem. In \Cref{SUB_paths}, we give an efficient $O(n^3)$ algorithm for the orientation problem when $\T$ is a path.

\subsection{Petals and stars}\label{SUB_petals_stars}

Given a vertex $x$, let $\mathcal{N}(x)$ be the set of its neighbors. A $x$-{\em petal} (or {\em petal} if there is no ambiguity) is a maximal subset $\mathcal{N}_i$ of $\mathcal{N}(x)$ such that, if $t, t' \in \N_i$, then, in any compatible orientation of $\T$, we have either $x\rightarrow t \land x \rightarrow t'$ or  $t \rightarrow x \land t'\rightarrow x$. Clearly, if two petals $\mathcal{N}_i$ and $\mathcal{N}_j$ are such that $\mathcal{N}_i\cap \mathcal{N}_j\neq\emptyset$, then $\mathcal{N}_i=\mathcal{N}_j$ and so the $x$-petals $\mathcal{N}_1, \ldots, \mathcal{N}_p$ form a partition of $\mathcal{N}(x)$.
This partition is computed by Algorithm~\ref{ALGO_petals_construction}.

\begin{proposition}
    Algorithm~\ref{ALGO_petals_construction} computes the petal partition of a vertex $x$ in $O(\deg^2(x))$.
\end{proposition}
\begin{proof}
    Since the distance $d(t,z)$ is tested for all $t,z \in \mathcal{N}(x)$, the partition $\N_1 \cup \N_2 \cup \ldots \cup \N_p$ does not depend on the choices that are made by the algorithm. In addition, as each distance is tested only once, Algorithm~\ref{ALGO_petals_construction} runs in $O(\deg^2(x))$.

    Clearly, in any compatible orientation of $T$, for any $t,t'$ in the same $x$-petal, we have either $x\rightarrow t \land x \rightarrow t'$ or  $t \rightarrow x \land t'\rightarrow x$. Now we show that, for  $t,t'$ neighbors of  $x$ but not in the same $x$-petal, there exists a compatible orientation of $T$ with $t\rightarrow x$ and $x\rightarrow t'$. 
    We orient the $x$-petal of $t$ towards $x$, the other $x$-petals from $x$, and, for all vertices $y\neq x$, we orient all its incident edges either from $y$ or towards $y$. This orientation is compatible since the only paths of length $>1$ are those having $x$ as a point in the middle.
 \end{proof}

\begin{algorithm2e}
  \caption{\textsc{Petals-Construction}}
  \label{ALGO_petals_construction}
  \KwIn{A dissimilarity space $(X, d)$; $x \in X$; a tree $T$ with vertex set $X$. $\mathcal{N}(x)$ is the set of neighbors of $x$.}
  \KwOut{A partition $\mathcal{N}_1 \cup \mathcal{N}_2 \cup \ldots$ of $\mathcal{N}(x)$.}
  \BlankLine
  $N \gets \mathcal{N}(x)$\;
  $i \gets 1$\;
  \While{$N \neq \emptyset$}{
    $\mathcal{N}_i \gets \{y\}$, where $y$ is any point in $N$\;
    \For{$z \in \mathcal{N}_i$}{
      \For{$t \in N \setminus \mathcal{N}_i$}{
        \If{$d(t,z) < \max\{d(x,t), d(x,z)\}$}{
          $\mathcal{N}_i \gets \mathcal{N}_i \cup \{t\}$\;
        }
      }
    }
    $N \gets N \setminus \mathcal{N}_i$\;
    $i \gets i + 1$\;
  }
  \Return{$\mathcal{N}_1, \mathcal{N}_2, \ldots$}
\end{algorithm2e}
\begin{comment}
    
\begin{algorithm}[htb]
  \caption{\textsc{Petals-Construction}
    \label{ALGO_petals_construction}}
    %{\small{
  \KwIn{
    A dissimilarity space $(X, d)$; $x \in X$; a tree $\T$ with vertex set $X$. $\mathcal{N}(x)$ is the set of neighbors of $x$.
  }
  \KwOut{
    A partition $\N_1 \cup \N_2 \cup \ldots $ of  $\N(x)$.
  }
  \Begin{
    $N \gets \N(x)$  \;
    $i \gets 1$ \;
    \While {$N \neq \emptyset$} {
        $\N_i \gets \{y\}$, where $y$ is any point in $N$ \;
        \ForAll{$z \in \N_i$}
            {
            \ForAll{$t \in N\setminus \N_i$}
                {\If {$d(t,z) < \max\{d(x,t), d(x, z)\}$}
                    {$\N_i \gets \N_i\cup \{t\}$ }
                }
            }
        $N \gets N \setminus \N_i$ \;
        $i \gets i + 1$ \;
        }
    \Return{$\N_1, \N_2,\ldots$}
    } %}}
\end{algorithm}
\end{comment}

We now give an $O(n^2)$ algorithm to compute an optimal orientation for the star $K_{1, n-1}$. We suppose that the center of the star is $x_1$.
% Algorithm~\ref{ALGO_petals_construction} {\sc Neighborhood-Partition} partitions the 

We first apply Algorithm~\ref{ALGO_petals_construction} to $K_{1, n-1}$ and $x_1$, and we get the $x_1$-petals $\mathcal{N}_1, \ldots, \mathcal{N}_p$.
To get an optimal orientation of $K_{1, n-1}$, we have to orient each $\N_i$ (into $\N_i \rightarrow x_1$ or $x_1 \rightarrow \N_i$) in a way that maximizes the number of paths. As for the constant dissimilarity, this corresponds to finding a set $I\subset \{1,\ldots, p\}$ such that $\sum_{i\in I}{\vert\N_i\vert}$ is the closest from $(n-1)/2$ (see \cref{PROP_best_orientation_with_fixed_central}).
This can be done by Algorithm~\ref{ALGO_optimal_neigborhood_partition} {\sc Optimal-Partition-of-Neighbors} by taking one $y_i$ in each $\N_i$ with $\theta(y_i) = \vert\N_i\vert$.
So we have:

\begin{proposition} \label{PROP_algo_orientation_star}
    Given a dissimilarity space $(X,d)$ with $\vert X \vert = n$ and a star $S$ with vertex set $X$, an optimal orientation of $S$ (compatible with $d$) can be found in optimal time $O(n^2)$.
\end{proposition}

We now consider the assignation problem. Given a vertex $v$ as the center of the star, the algorithm for the assignation problem is the same as for the orientation one. As we need to check for every vertex as the center of the path, we get an $O(n^3)$ algorithm.

\subsection{Paths}\label{SUB_paths}

We now consider the path $(x_1, x_2, \ldots, x_n)$. We first determine, for each $1\leq i <n$, the greatest $\eta(i)$ such that $x_i\rightarrow x_{i+1}\rightarrow \ldots \rightarrow x_{\eta(i)}$ is Robinson.
Notice that, if $i< i'$, then $\eta(i) \leq \eta(i')$.

This is done by  Algorithm~\ref{ALGO_eta_computation} $\eta$-{\sc Computation}:

\begin{proposition}
    If $(X, d)$, with $X = \{x_1,x_2,\ldots, x_n\}$ is a dissimilarity space, then after Algorithm~\ref{ALGO_eta_computation} $\eta$-{\sc Computation}, the resulting sequence $[(x_{i_k}, x_{j_k}), 1\leq k \leq p]$ is such that:
    \begin{enumerate}
    \item $ {i_1} = 1$, $j_k = n$,

    \item  $\forall\, 1\leq k \leq p$, $x_{j_k} = \eta(x_{i_k})$,

    \item  $\forall 1\leq k < p$, $\forall i_k \leq i < i_{k+1}$, $\eta(x_i) = \eta(x_{i_k})$.
    \end{enumerate}
\end{proposition}

\begin{proof}
    The Loop at Line~\ref{LINE_3_7} searches the smallest $k$ such that $x_k \rightarrow \ldots \rightarrow x_j$ is a compatible orientation.
    If $k = i$, then we check $j+1$ (Line~\ref{LINE_3_12}).
    If $k > i$, then $x_i\rightarrow\ldots\rightarrow x_{j-1}$ is a compatible orientation, but not $x_i\rightarrow\ldots\rightarrow x_j$, and this is also the case with $x_{i'}$ instead of $x_i$ for all $i\leq i' < k$. So, for all $i\leq i'< k$, $\eta(x_{i'}) = x_{j-1}$ and Conditions 2 and 3 are satisfied.
    Condition 1 is satisfied by Lines~\ref{LINE_3_1} and \ref{LINE_3_n}.
\end{proof}

\begin{proposition}
    Algorithm~\ref{ALGO_eta_computation} $\eta$-{\sc Computation} runs in $O(n^2)$.
\end{proposition}
\begin{proof}
    Apart from the two loops, operations in Algorithm~\ref{ALGO_eta_computation} take a constant time.
\end{proof}
We now use the values of $\eta$ to determine the optimal orientation of the path $(x_1, \ldots, x_n)$.
We will for that construct two $n\times n$ matrices $\M$ and $\MM$ where, for $i<j$:
\begin{itemize}
    \item 
$\M[i, j]$ is the number of compatible paths in the optimal orientation of $x_i,\ldots, x_j$.
Notice that, if $i<j\leq \eta(i)$, the optimal orientations of $x_i\ldots x_j$ are $x_i\rightarrow \ldots \rightarrow x_j$ and its reverse, and so $\M[i, j] = (j-i+1)(j-i)/2$. 
    \item 
    If $\MM[i,j] = 0$, then $x_i\rightarrow \ldots\rightarrow x_j$ is an optimal orientation of $(x_i,\ldots x_j)$. If that is not the case,
    there exists an optimal orientation of $(x_i,\ldots x_j)$ with $x_{\MM[i,j]-1} \not\leftrightsquigarrow x_{\MM[i,j]+1}$. 
\end{itemize}
Matrices $\M$ and $\MM$ are computed by Algorithm~\ref{ALGO_path_orientation} {\sc Path-Orientation}, which follows the classical dynamic programming paradigm and runs in $O(n^3)$.

After Algorithm~\ref{ALGO_path_orientation} {\sc Path-Orientation}, for all $i<j$, if $\MM[i,j] = 0$, there exists a compatible orientation $x_i\rightarrow \ldots\rightarrow x_j$; otherwise, there is an optimal orientation of $(x_i,\ldots, x_j)$ which is made of an optimal orientation of $(x_i,\ldots, x_{\MM[i,j]})$ and of an optimal orientation of $(x_{\MM[i,j]}, \ldots ,x_j)$.
Starting from $\MM[1,n]$, it is thus possible to build an optimal orientation of $(x_1,\ldots, x_n)$ in $O(n)$.
So we have:

\begin{algorithm2e}
  \caption{$\eta$-\textsc{Computation}}
  \label{ALGO_eta_computation}
  \KwIn{A dissimilarity space $(X, d)$ with $X = \{x_1, x_2, \ldots, x_n\}$. Implicitly, $(x_1, x_2, \ldots, x_n)$ is a path.}
  \KwOut{A sequence of pairs $(x_i, \eta(x_i))$.}
  \BlankLine
  $Resulting\_Sequence \gets []$\;
  $i \gets 1$\; \label{LINE_3_1}
  $j \gets 3$\;
  \While{$j \leq n$}{
    $k \gets j$\;
    \While{$k > i \,\land\, d(x_j, x_k) \leq d(x_j, x_{k-1}) \,\land\, d(x_{j-1}, x_{k-1}) \leq d(x_j, x_{k-1})$}{
      $k \gets k - 1$\; \label{LINE_3_7}
    }
    \If{$k > i$}{
      $Resulting\_Sequence \gets Resulting\_Sequence + [(x_i, x_{j-1})]$\;
      $i \gets k$\;
    }
    $j \gets j + 1$\; \label{LINE_3_12}
  }
  $Resulting\_Sequence \gets Resulting\_Sequence + [(x_i, x_n)]$\; \label{LINE_3_n}
  \Return $Resulting\_Sequence$\;
\end{algorithm2e}

\begin{comment}
\begin{algorithm}[ht]
  \caption{$\eta$-\textsc{Computation}
    \label{ALGO_eta_computation}}
    %{\small{
  \KwIn{
    A dissimilarity space $(X, d)$ with $X = \{x_1, x_2, \ldots, x_n\}$.
    Implicitly, $(x_1,x_2,\ldots, x_n)$ is a path.
  }
  \KwOut{
    A sequence of couples $(x_i, \eta(x_i))$.
  }
  \Begin{
    $Resulting\_Sequence\, \gets []$ \;
    $i\, \gets 1$ \; \label{LINE_3_1}
    $j \,\gets 3$ \;
    \While {j $\leq n$}
        {$ k \gets j$ \;
        \While {$k> i \,\land\, d(x_j, x_k) \leq d(x_j, x_{k-1}) \,\land\, d(x_{j - 1}, x_{k-1}) \leq d(x_j, x_{k-1})$} 
            {
            $k \gets k -1$ \label{LINE_3_7}
            }
        \If {$k > i$}
            {$Resulting\_Sequence \gets Resulting\_Sequence + [(x_i, x_{j-1})]$ \;
            $ i \gets k $
            }
        $j \gets j + 1$ \label{LINE_3_12}
        }
    $Resulting\_Sequence \gets Resulting\_Sequence + [(x_i, x_{n})]$ \; \label{LINE_3_n}
    \Return $Resulting\_Sequence$
    }
\end{algorithm}
\end{comment}
%
\begin{algorithm2e}
  \caption{\textsc{Path-Orientation}}
  \label{ALGO_path_orientation}
  \KwIn{A dissimilarity space $(X, d)$ with $X = \{x_1, x_2, \ldots, x_n\}$ (implicitly, $(x_1, x_2, \ldots, x_n)$ is a path). Each $x_i$ is given with $\eta(x_i)$.}
  \KwOut{Two $n \times n$ matrices $\M$ and $\MM$.}
  \BlankLine
  \ForAll{$1 \leq i \leq n$}{
    \ForAll{$i \leq j \leq n$}{
      \eIf{$i < j \,\land\, x_j \leq \eta(x_i)$}{
        $\M[i,j] \gets (j-i+1)(j-i)/2$\;
      }{
        $\M[i,j] \gets 0$\;
      }
      $\MM[i,j] \gets 0$\;
    }
  }
  \ForAll{$1 \leq \kappa < n$}{
    \ForAll{$1 \leq i \leq n-\kappa$}{
      $j \gets i + \kappa$\;
      \ForAll{$i < k < j$}{
        \label{LINE_4_12}
        \If{$\M[i,j] < \M[i,k] + \M[k,j]$}{
          $\M[i,j] \gets \M[i,k] + \M[k,j]$\;
          $\MM[i,j] \gets k$\;
        }
      }
    }
  }
\end{algorithm2e}

\begin{comment}
\begin{algorithm}[ht]
  \caption{\textsc{Path-Orientation}
    \label{ALGO_path_orientation}}
    %{\small{
  \KwIn{
    A dissimilarity space $(X, d)$ with $X = \{x_1, x_2, \ldots, x_n\}$ (implicitly, $(x_1,x_2,\ldots, x_n)$ is a path). Each $x_i$ is given with $\eta(x_i)$.
  }
  \KwOut{
    Two $n\times n$ matrices $\M$ and $\MM$.
  }
  \Begin{
    \ForAll{$1\leq i \leq n$}
        {\ForAll{$i \leq j \leq n$}
            {
            \eIf{$i < j \,\land\, x_j \leq \eta(x_i)$}
                {$\M[i,j]\gets (j-i+1)(j-i)/2$}
                {$\M[i,j] \gets 0$}
            $\MM[i,j] \gets 0$
            }
        }
    \ForAll{$1\leq \kappa < n$}
        {\ForAll{$1\leq i\leq n-\kappa$}
            {
            $j \gets i + \kappa$ \;
            \ForAll{$i< k < j$}
                {
                \label{LINE_4_12}
                \If{$\M[i,j] < \M[i,k] + \M[k,j])$}
                    {$\M[i,j] \gets  \M[i,k] + \M[k,j]$ \;
                    $\MM[i,j] \gets k$}
                }
            }
        }
    }
\end{algorithm}
\end{comment}
%
\begin{proposition}\label{PROP_algo_orientation_path}
    Given a dissimilarity space $(X, d)$ with $X= \{x_1,\ldots, x_n\}$, it is possible to construct an optimal orientation of the path $(x_1, \ldots, x_n)$ in $O(n^3)$.
\end{proposition}
The $O(n^3)$ complexity is entirely due to Algorithm~\ref{ALGO_path_orientation} (the other steps are in $O(n^2)$ and $O(n)$). 
It is possible to improve Algorithm~\ref{ALGO_path_orientation} by considering, in the loop at Line~\ref{LINE_4_12}, only the $k$'s such that $\eta(x_{k-1}) \neq \eta(x_k)$ or such that there exists $k'$ with $\eta(x_{k'}) = x_k$. 
These $k$'s are the indices returned by Algorithm~\ref{ALGO_eta_computation} $\eta$-{\sc Computation} and Algorithm~\ref{ALGO_path_orientation} would then run in $O(p^3)$, where $p$ is the length of the sequence returned by Algorithm~\ref{ALGO_eta_computation}.
This does not change the worst-case complexity of the algorithm.

\section{Conclusion}\label{SECTION_conclusion}

In this work, we extended the concept of Robinson spaces to include asymmetric dissimilarities, significantly broadening their scope and applicability. Within this extended framework, we introduced two new problems: the Assignment problem and the Orientation problem, both generalizing the classical seriation problem. We proved that the Assignment problem is NP-complete and the Orientation problem is NP-hard, highlighting the computational challenges inherent to these formulations. Nevertheless, we identified several non-trivial cases where these problems can be solved efficiently in polynomial time, providing valuable insights into their structure. These results deepen our understanding of asymmetric Robinson spaces.

Regarding future work, we believe exploring the complexity of specific cases warrants further investigation. In particular, the decision problem for one-way-Robinson spaces and the Assignment and Orientation problems when the input structure is a star with non-symmetric dissimilarities present intriguing challenges. 

%\section*{Acknowledgments}
%We would like to acknowledge the assistance of volunteers in putting
%together with this example manuscript and supplement.

\end{document}